\documentclass[runningheads]{llncs}

\usepackage{hyperref}
\usepackage{graphicx}
\usepackage{csvsimple}
\usepackage{comment}
\usepackage{multirow}
\usepackage[table]{xcolor}
\usepackage{xcolor}
\usepackage[]{algorithm2e}
\usepackage{url}
\usepackage{multirow}
\usepackage{amsmath}
\usepackage{rotating}
\usepackage[T1]{fontenc}
\usepackage{soul}
\usepackage{tikz}
 
\newcommand{\furl}[1]{\footnote{\scriptsize \url{#1}}}
\colorlet{punct}{red!60!black}
\definecolor{background}{HTML}{EEEEEE}
\definecolor{delim}{RGB}{20,105,176}
\colorlet{numb}{magenta!60!black}
\usepackage{array}

\begin{document}

\title{Extracting Topics from Open Educational Resources}

\author{Mohammadreza Molavi\inst{1} \and
Mohammadreza Tavakoli\inst{2} \and
G\'abor Kismih\'ok\inst{2}}
\authorrunning{M. Molavi et al.}

\institute{Amirkabir University of Technology, Tehran, Iran \\
\email{mr.molavi@aut.ac.ir}  \\
\and German National Library of Science and Technology (TIB), Hannover, Germany\\
\email{\{reza.tavakoli,gabor.kismihok\}@tib.eu}}

\maketitle

\begin{abstract}
In recent years, Open Educational Resources (OERs) were earmarked as critical when mitigating the increasing need for education globally. Obviously, OERs have high-potential to satisfy learners in many different circumstances, as they are available in a wide range of contexts. However, the low-quality of OER metadata, in general, is one of the main reasons behind the lack of personalised services such as search and recommendation. As a result, the applicability of OERs remains limited. Nevertheless, OER metadata about covered topics (subjects) is essentially required by learners to build effective learning pathways towards their individual learning objectives. Therefore, in this paper, we report on a work in progress project proposing an OER topic extraction approach, applying text mining techniques, to generate high-quality OER metadata about topic distribution. This is done by: 1) collecting 123 lectures from \emph{Coursera} and \emph{Khan Academy} in the area of data science related skills, 2) applying \emph{Latent Dirichlet Allocation} (\emph{LDA}) on the collected resources in order to extract existing topics related to these skills, and 3) defining topic distributions covered by a particular OER. To evaluate our model, we used the data-set of educational resources from \emph{Youtube}, and compared our topic distribution results with their manually defined target topics with the help of \emph{3} experts in the area of data science. As a result, our model extracted topics with \textbf{79\%} of \emph{F1-score}. 

\end{abstract}

\section{Introduction}
The world of education is changing rapidly due to the growing needs of personalized services. The quickly evolving labour market, its dynamically changing knowledge and skills demands \cite{tavakoli2020labour}, the global challenges for work and education due to the emergence of the COVID-19 pandemic, are all examples that highlight the increasing need for flexible and personalised education. Consequently, we have been facing with an exponential growth of distributed and heterogeneous educational materials (such as Open Educational Resources (OERs)) \cite{saraiva2019relationships,de2017finding}. Enormous amount of OERs are provided and receiving more and more attention from learners every day. However, OER authors often fail to provide metadata for their content, as they consider this as a time-consuming activity \cite{garcia2017social}. This notion leads to low-quality OER metadata, even though that high quality metadata is crucial for organizing OERs \cite{garcia2017social} and providing search and recommendation services \cite{tavakoli2020quality}. Indeed, the lack of high quality metadata is one of the most important factors limiting the effectiveness of (OER based) personalised informal learning \cite{wang2015topic}. The \emph{Covered Topics} (i.e. covered knowledge areas) is one of the essential (often ignored) metadata for learning resources, as it helps learners to find the most suitable educational resource for their learning objectives \cite{saraiva2019relationships,de2017finding}. Having a clear picture about \emph{Covered Topics} is especially important for OER users, who want to build their own learning trajectory \cite{wang2015topic}.
As a potential solution, intelligent algorithms should be tailored to extract and identify topics from OERs automatically, which emerged as a key issue in e-learning in the recent decade \cite{xie2011new}. To tackle this issue, researchers have elaborated a number of methods already. For instance, they have applied machine learning methods on educational materials \cite{saraiva2019relationships,de2017finding,garcia2017social}, or analyzed and transformed educational content into appropriate data structures (e.g. trees)\cite{wang2015topic} to extract a particular topic from educational resources. However, the generalisability of these proposed approaches, so far, has remained limited \cite{saraiva2019relationships}.   

In this paper, we address the above mentioned challenges and propose an automatic topic extraction model focusing on data science related OERs. This is done by 1) collecting OERs about data science related skills (e.g. machine learning, text mining, sql language) from two pioneer educational repositories (\emph{Coursera, Khan Academy}), 2) identifying topics that should be ideally covered in each skill using topic modeling techniques (i.e. \emph{LDA}), and 3) building topic models to extract the distribution of \emph{Covered Topics} for a given OER. We evaluated our model by using a data-set of open educational videos from Youtube, assigning their topics (as labels) with the help of \emph{3} data science experts, and calculating \emph{F1-score} (harmonic mean of precision and recall) of our automatic topic distribution extractor, using manually assigned topics as ground truth.


\section{State of the Art}

\subsection{Semantic-based Methods}
A number of studies use semantic methods and structured representation of data (such as taxonomies) to extract topics from educational resources~\cite{de2017finding}. For instance, \cite{saraiva2019relationships} proposed a framework that combines semantic classification, taxonomies, and graph structures to extract topics and detect their relationships.

\subsection{Text Mining Methods}
Studies in this group analyse educational text and use text-related machine learning methods to detect topics in educational text~\cite{xie2011new}.
For example, \cite{wang2015topic} created a system, which collects domain-specific content from online learning systems. They extracted domain-specific terms by creating \emph{Generalised Suffix Tree} (\emph{GST}) from resources' text, and detected repeated sub-sequences as candidate terms to provide topic-specific recommendations for learners.

\section{Method}
\subsection{Data Collection and Pre-processing}
\subsubsection{Target Skills}
To propose the first version of our approach, we extracted important skills for data science by mining relevant job vacancies \cite[page\_8]{tavakoli2020labour}. In our online job vacancy data-set (from August 2019 to December 2019) the three most important data science skills were 1. \emph{Machine Learning}, 2. \emph{Text Mining}, and 3. \emph{SQL Language}. 

\subsubsection{OER Resources}
In order to build our topic models, we collected 123 relevant online lectures (and their transcripts) from \emph{Coursera}\footnote{\url{https://www.coursera.org/}} and \emph{Khan Academy}\footnote{\url{https://www.khanacademy.org/}} related to our target skills (including 67 lectures for machine learning, 27 for text mining, and 29 for sql language).
Moreover, to evaluate our proposed model, we used the data-set defined by \cite[page\_3]{tavakoli2020recommender} including 550 educational videos and their properties (e.g. rate, transcript, view-count) from \emph{Youtube} in the area of data science. It should be mentioned that we applied the following pre-processing steps to prepare our collected OER transcripts for our analysis: 1) Removal of unimportant characters, punctuations, links, and stop words, and 2) Building TF-IDF representation.

\subsection{Building Topic Models}
To extract knowledge areas that are covered by particular educational resources in each of the target skills, we used \emph{Latent Dirichlet Allocation} (\emph{LDA}) \cite{jelodar2019latent}. \emph{LDA} is a generative probabilistic topic model that considers each document as a distribution of different topics, each topic as a distribution of different words, and tries to extract existing topics together with their distribution of words for a corpus. To set the number of topics that \emph{LDA} extracts, we calculated $C_V$ \emph{Coherence} \cite{roder2015exploring} for different number of topics (between 2 to 10), and selected the topic amount with the highest coherence value. The following part of this section explains the detailed process of finding the most appropriate value of topic amounts, extracting topics, and assigning a name to topics (done manually, after executing \emph{LDA} and based on the distribution of their words \cite{blei2003latent}) for \emph{Text Mining} skill\footnote{We report our result regarding topic amounts and evaluation for the other two skills: \emph{Machine Learning} and \emph{SQL Language}}.

\begin{figure*}[!htbp]
    \centering
    \includegraphics[width=0.6\textwidth, height=5cm]{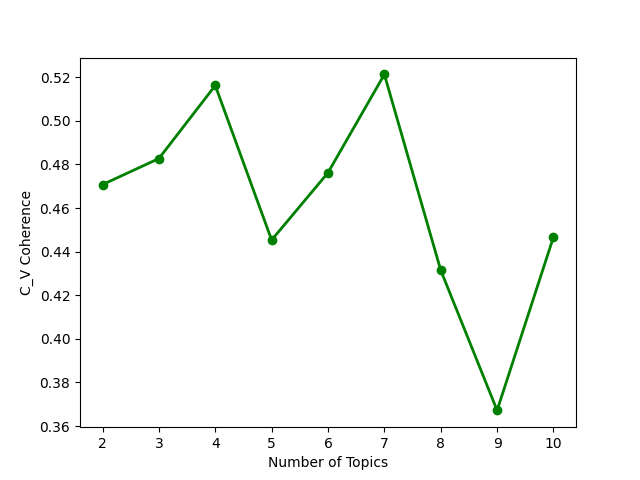}
    \vspace{-0.4cm}
    \caption{$C_V$ coherence for different number of topics in Text Mining corpus}
    \label{fig:text}
\end{figure*}

In order to build our topic models on \emph{Text Mining} skill, we calculated $C_V$ coherence for different number of topics on text mining-related educational resources, as shown on Figure~\ref{fig:text}. Based on the result, $7$ topics provide us the best coherence value\footnote{We did the same process and set $C_V$ of other skills as follows: \emph{Machine Learning}: 9, \emph{SQL Language: 5}}. Therefore, we set the parameter \emph{k} of \emph{LDA} to $7$ and executed the analysis on our text mining corpus. The extracted topics, the assigned names, and the significant words of each topic are visible on Table~\ref{tab:text}.

\vspace{-0.7cm}

\begin{table*}[!htbp]
\vspace{-0.9cm}
 \caption{Output of LDA on Text Mining corpus}
\begin{center}
\vspace{-0.6cm}
\resizebox{\textwidth}{!}{
\begin{tabular}{|c|c|c|}
\hline
\rowcolor[HTML]{C0C0C0}
\textbf{Topics} & \textbf{Assigned Name} & \textbf{Significant Words}\\ \hline
$Topic_{1}$ & Topic Modeling & lda plsa topic dirichlet parameters likelihood beta distribution alpha document \\ \hline
$Topic_{2}$ & Sequence Models & prior string tag sequence markov hidden probabilities estimate position generate \\ \hline
$Topic_{3}$ & Sentiment Analysis & features grams sentiment positive topic accuracy reviews negative idf tf \\ \hline
$Topic_{4}$ & Matrix Factorization & matrix topic matrices squared diagonal svd factorization vectors approximation document \\ \hline
$Topic_{5}$ & Text Classification & grams convolutional filter sentence corpus vec embeddings neural count modeling  \\ \hline
$Topic_{6}$ & Probabilistic Models & naive prior bayes given probability likelihood independent maximizes predicted significantly \\ \hline
$Topic_{7}$ & \shortstack{Text Process \&\\Feature Extraction} & sentence document frequency features phrase vec grammar parse grams term \\ \hline
\end{tabular}}
\label{tab:text}
\end{center}
\end{table*} 

\vspace{-1cm}
\section{Topic Model Extraction Evaluation}
\vspace{-0.4cm}
To evaluate our topic models, we used our Youtube data-set in which topics were assigned to videos manually. This manual assignment was done by 3 data science experts with at least 2 years of teaching experience and 5 years of industrial experience in data science related areas. It should be mentioned that each participant allocated at least 2 minutes for analysing each of the videos. Afterwards, we applied our topic extraction models on each video transcript (e.g. apply our machine learning topic model on the related educational videos on machine learning). Finally, we compared the manually assigned topics (by experts) and the output of our topic extraction models. As a result, we were able to determine the quality of our topic extraction models in relation to manual, expert topic assignments. We got \emph{F1-score} of each topic extraction model as follows: \emph{Text Mining}: \emph{76\%}, \emph{Machine Learning}: \emph{81\%}, and \emph{SQL Language}: \emph{78\%}. Therefore, our models were able to extract covered topics of educational resources with F1-score of \textbf{79\%} in average.

\section{Conclusion and Future Work}
\vspace{-0.1cm}
This study is one of the steps towards 1) dynamic definition of topics that should be covered in a particular knowledge areas, and 2) extracting the topic distribution for a given OER, as one of the most important metadata, to help learners to build their own learning path. We collected 123 educational lectures from two repositories related to 3 data science related skills. After that we applied \emph{LDA} on the lectures' transcripts to extract the topic model for each skill. Finally, to evaluate the models, we used an educational Youtube data-set, assigned covered topics with the help of 3 data science experts. Subsequently, we applied our topic extraction models, and compared the output of our model with the manually assigned topics. This exercise revealed that our models can extract topics with \emph{F1-score} of \textbf{79\%}. As the next steps, we plan to add more educational resources to improve our models and also, apply our approach for other skills/knowledge.

\vspace{-0.11cm}
\bibliographystyle{splncs04}
\bibliography{paper}

\begin{thebibliography}{10}
\providecommand{\url}[1]{\texttt{#1}}
\providecommand{\urlprefix}{URL }
\providecommand{\doi}[1]{https://doi.org/#1}

\bibitem{blei2003latent}
Blei, D.M., Ng, A.Y., Jordan, M.I.: Latent dirichlet allocation. Journal of
  Machine Learning Research  \textbf{3}(Jan),  993--1022 (2003)

\bibitem{de2017finding}
de~Carvalho~Saraiva, M., Medeiros, C.B.: Finding out topics in educational
  materials using their components. In: 2017 IEEE Frontiers in Education
  Conference (FIE). pp.~1--7. IEEE (2017)

\bibitem{garcia2017social}
Garc{\'\i}a-Floriano, A., Ferreira-Santiago, A., Y{\'a}{\~n}ez-M{\'a}rquez, C.,
  Camacho-Nieto, O., Aldape-P{\'e}rez, M., Villuendas-Rey, Y.: Social web
  content enhancement in a distance learning environment: intelligent metadata
  generation for resources. International Review of Research in Open and
  Distributed Learning  \textbf{18}(1),  161--176 (2017)

\bibitem{jelodar2019latent}
Jelodar, H., Wang, Y., Yuan, C., Feng, X., Jiang, X., Li, Y., Zhao, L.: Latent
  dirichlet allocation (lda) and topic modeling: models, applications, a
  survey. Multimedia Tools and Applications  \textbf{78}(11),  15169--15211
  (2019)

\bibitem{roder2015exploring}
R{\"o}der, M., Both, A., Hinneburg, A.: Exploring the space of topic coherence
  measures. In: Proceedings of the eighth ACM international conference on Web
  search and data mining. pp. 399--408 (2015)

\bibitem{saraiva2019relationships}
Saraiva, M.d.C., et~al.: Relationships among educational materials through the
  extraction of implicit topics: Relacionamentos entre materiais did{\'a}ticos
  atrav{\'e}s da extra{\c{c}}{\~a}o de t{\'o}picos impl{\'\i}citos  (2019)

\bibitem{tavakoli2020quality}
Tavakoli, M., Elias, M., Kismih{\'o}k, G., Auer, S.: Quality prediction of open
  educational resources a metadata-based approach. arXiv preprint
  arXiv:2005.10542  (2020)

\bibitem{tavakoli2020recommender}
Tavakoli, M., Hakimov, S., Ewerth, R., Kismih{\'o}k, G.: A recommender system
  for open educational videos based on skill requirements. arXiv preprint
  arXiv:2005.10595  (2020)

\bibitem{tavakoli2020labour}
Tavakoli, M., Mol, S.T., Kismih{\'o}k, G.: Labour market information driven,
  personalized, oer recommendation system for lifelong learners. arXiv preprint
  arXiv:2005.07465  (2020)

\bibitem{wang2015topic}
Wang, J., Xiang, J., Uchino, K.: Topic-specific recommendation for open
  education resources. In: International Conference on Web-Based Learning. pp.
  71--81. Springer (2015)

\bibitem{xie2011new}
Xie, M., Wu, C., Zhang, Y.: A new intelligent topic extraction model on web.
  JCP  \textbf{6}(3),  466--473 (2011)

\end{thebibliography}

\end{document}